# VIBRATION-INDUCED CONDUCTIVITY FLUCTUATION (VICOF) TESTING OF SOILS [*]

L. B. KISH[†],

*Department of Electrical and Computer Engineering, Texas A&M University, College Station, TX 77843-3128, USA*

C. L. S. MORGAN, and A. SZ. KISHNÉ

*Department of Soil and Crop Science, Texas A&M University, College Station, TX 77843-2474, USA*

In this Letter, we propose and experimentally demonstrate a simple method to provide additional information by conductivity measurements of soils. The AC electrical conductance of the soil is measured while it is exposed to a periodic vibration. The vibration-induced density fluctuation implies a corresponding conductivity fluctuation that can be seen as combination frequency components, the sum and the difference of the mean AC frequency and the double of vibration frequency, in the current response. The method is demonstrated by measurements on clayey and sandy soils.

*Keywords*: soil water content; salinity; soil bulk density; soil connectivity; soil electrical conductivity; conductivity fluctuations.

## 1. The new measurement principle

The bulk electrical conductivity of soils depends on various soil properties, such as water content, salt type and concentration, bulk density (air-filled porosity), clay content and mineralogy, and connectivity structure of soil particles [1]. Therefore, given electrical conductivity measurement data can be the result of many different combinations of soil properties. The interaction of soil properties complicates interpretation of soil electrical conductivity measurements. Soil electrical conductivity sensors such as capacitance sensors [2-4], electromagnetic induction [5-6], and resistivity tomography [7] are used to quantify soil moisture, salinity, clay content, water flux, and other related soil properties; however, the empirical calibrations used for these sensors are very site specific and therefore limited in their application. The goal of this paper is to propose and demonstrate a new technique based on vibration-induced modulation of the electrical conductivity that gives additional and independent information about the *mechano-electrical transport properties* of the soil. With proper models, these electrical transport properties can provide more information about the soil structure, such as soil porosity, and the associated empirical calibrations would be more robust.

The measurement circuitry, which is an expanded version of the standard AC conductivity measurement circuitry, is shown in Figure 1. The AC voltage generator provides a sinusoidal voltage at the main frequency $f_1$ that drives an AC current through the driving resistor $R_1$ and the resistor $R_s$ represented by the soil sample. The soil sample is

---

[*] The content of this paper is the subject of a Texas A&M University patent disclosure submitted on 10/31/05.
[†] Until 1999, L.B. Kiss, corresponding author

*Vibration-induced conductivity fluctuation in soils*

exposed to a *weak* periodic vibration (shaking) with frequency $f_2$. This implies a periodic pressure and density modulation at frequency $2f_2$, inducing a conductance modulation (similarly to a *carbon microphone*) with $2f_2$ first harmonics and that yields voltage components at the combination frequencies $f_1 + 2f_2$ and $f_1 - 2f_2$.

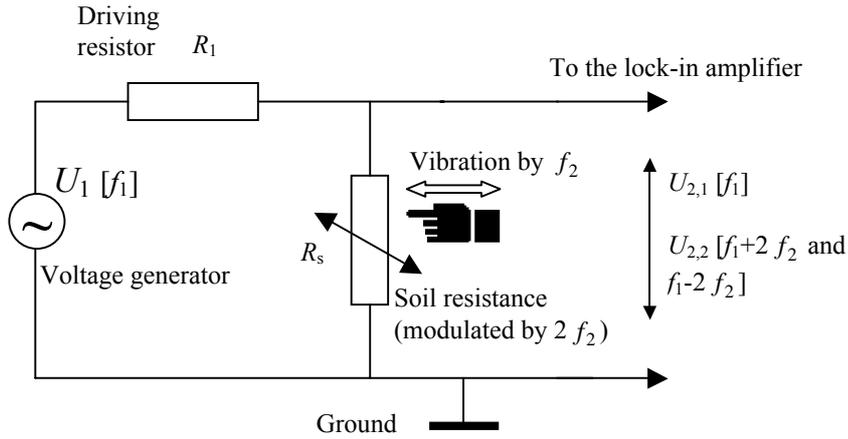

**Figure 1.** The measurement circuitry. The soil sample is exposed to a periodic vibration with frequency $f_2$. This implies a periodic pressure and density modulation inducing a conductance modulation with $2f_2$ first harmonics and that yields voltage components at the combination frequencies $f_1+2f_2$ and $f_1-2f_2$ during the execution of an AC conductivity measurement with sinusoidal voltage of frequency $f_1$.

At small and sinusoidal vibration and corresponding linear response, the voltage on the soil resistance has three frequency components. At the frequency of the voltage generator (main frequency, $f_1$) there is a classical AC conductance measurement response (voltage divider response):

$$U_{2,1} = U_1 \frac{R_s}{R_1 + R_s} , \tag{1}$$

This allows a determination of the AC resistance of the soil sample from the measurement of $U_{2,1}$ in the classical way:

$$R_s = R_1 \frac{U_{2,1}}{U_1 - U_{2,1}} . \tag{2}$$

Supposing small modulation, we can estimate the sensitivity of the amplitude $U_{2,1}$ against the modulation of the soil resistance as follows:

$$\frac{dU_{2,1}}{dR_s} = \frac{U_1 R_1}{(R_1 + R_s)^2} . \tag{3}$$

According to the following relation of *amplitude modulation*,

$$\sin(2\pi f_1 t)\sin(4\pi f_2 t) = \frac{1}{2}\cos[2\pi(f_1 + 2f_2)t] + \frac{1}{2}\cos[2\pi(f_1 - 2f_2)t],$$

it can be seen that the modulation yields the following amplitude components at the combination frequencies $f_1 + 2f_2$ and $f_1 - 2f_2$:

$$U_{2,2} = U_{2,2}[f_1 + 2f_2] = U_{2,2}[f_1 - 2f_2] = \frac{1}{2}dU_{2,1} = \frac{1}{2}\frac{U_1 R_1}{(R_1 + R_s)^2}dR_s, \quad (4)$$

where $U_{2,2}$ is the signal above the average background voltage at the combination frequencies. From Eqs. 1, 2 and 4:

$$dR_s = 2\frac{U_{2,2}}{U_1}\frac{(R_1 + R_s)^2}{R_1} = 2\frac{U_{2,2}}{U_1 - U_{2,1}}(R_1 + R_s) = 2\frac{U_{2,2}}{U_1 - U_{2,1}}(R_1 + R_1 \frac{U_{2,1}}{U_1 - U_{2,1}}), \quad (5)$$

and the *fluctuation amplitude* of the soil resistance can be determined from the known driving resistance $R_1$ and the measurement of the AC voltage amplitudes:

$$dR_s = 2R_1 \frac{U_{2,2}}{U_1 - U_{2,1}}(1 + \frac{U_{2,1}}{U_1 - U_{2,1}}). \quad (6)$$

The normalized (relative) resistance fluctuation is especially important because it is probing the strength of modulation of the electrical connectivity properties of the soil. Its value can easily be determined from the above equations:

$$\frac{dR_s}{R_s} = 2R_1 \frac{U_{2,2}}{U_1 - U_{2,1}}\left(1 + \frac{U_{2,1}}{U_1 - U_{2,1}}\right)\left(R_1 \frac{U_{2,1}}{U_1 - U_{2,1}}\right)^{-1}, \quad (7)$$

and even the driving resistance is absent from this final form:

$$\frac{dR_s}{R_s} = 2\frac{U_{2,2}}{U_{2,1}}\left(1 + \frac{U_{2,1}}{U_1 - U_{2,1}}\right). \quad (8)$$

To evaluate the relative fluctuations of the soil resistance due to vibrations, we only need to know the above voltage components at relevant frequencies and use Eq. 8.

## 2. Experimental demonstration

The test experiments were carried out on an antivibration table, (100BM-2 Nano-K vibration isolation platform). An induction coil based vibrator (5W, 60Hz) was fixed to one side of the floating top of the antivibration table so that the vibration was horizontal in a well-defined direction. The soil sample contained in a tin sample holder (9.7 cm diameter and 6.3 cm height) was placed on this floating top. The ground contact was the metal container and the probing contact was provided by a standard cylindrical stainless steel electrode (3 mm diameter and 71 mm length). The electrode was placed 5 cm deep into the soil and 3 cm (*d*) from the wall of the tin. Figure 2 shows the top view of the arrangement. The lock-in amplifier was Stanford Research Systems SR830 DSP.

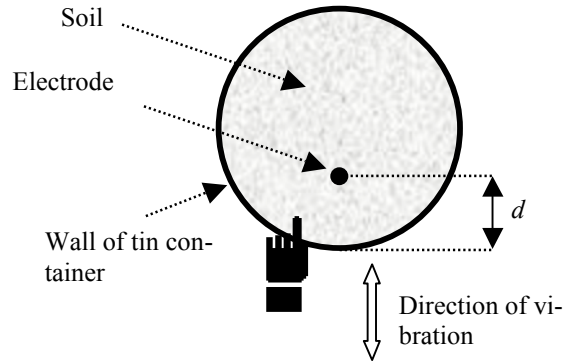

**Figure 2.** The arrangement of the electrode, soil sample, and vibration direction used for measurement.

The vibration was very weak, thus it did not cause any observable relaxation of the soil structure (compaction) which was concluded from the stable value of measured conductance. The voltage components at the combination frequency (1.12 kHz) were about $10^5$ times smaller than the amplitude at the main frequency ($f_1$=1 kHz).

The soil samples were non-saline and had clayey and fine sandy texture (Table 1). Soils were wetted by adding water to a pre-determined moisture content based on two matric potentials, -100 and -1000 J kg$^{-1}$. The two matric potentials were chosen to secure the comparable level of loosely held mobile water content and surface tension in the different soils. The amount of water was determined based on a relationship of gravimetric water content to matric potential as a function of soil texture [8]. To wet the soils, distilled and deionized water was added to air dry soils, mixed thoroughly by shaking in a sealed plastic bag, and allowed to equilibrate under constant temperature for 2 months. Three replicates of three compaction levels of each soil texture were measured, 36 samples totally. The soils were uniformly packed in 0.5-1 cm thick layers dropping a 1 kg weight from about a 2 cm. Soil surface was scratched between layers to ensure good contact with the next layer. After the measurements, soil samples were oven dried at 105 $^{o}$C to a constant weight and weighed again for moisture determination.

| Soil | Particle size distribution (mm) | | | Texture class | $EC_e^*$ |
|---|---|---|---|---|---|
| | Sand (2.0-0.05) | Silt (0.05-0.002) | Clay (<0.002) | | |
| | ----------------------------------%---------------------------------- | | | | dS m$^{-1}$ |
| 1 | 11.2 | 36.7 | 52.1 | clay | 0.6 |
| 2 | 97.6 | 0.9 | 1.5 | fine sand | 0.3 |

\* $EC_e$ = electrical conductivity of saturated paste extract

Table 1. Particle size distribution and electrical conductivity of soil samples.

Measurements of $U_1$ [1 kHz], $U_{2,1}$ [1 kHz], $U_{2,2}$ [1.12 kHz] with and without vibration (background) were taken with the electrode placed at three locations in 3 cm from the wall. Minimum three readings were averaged for each measurement.

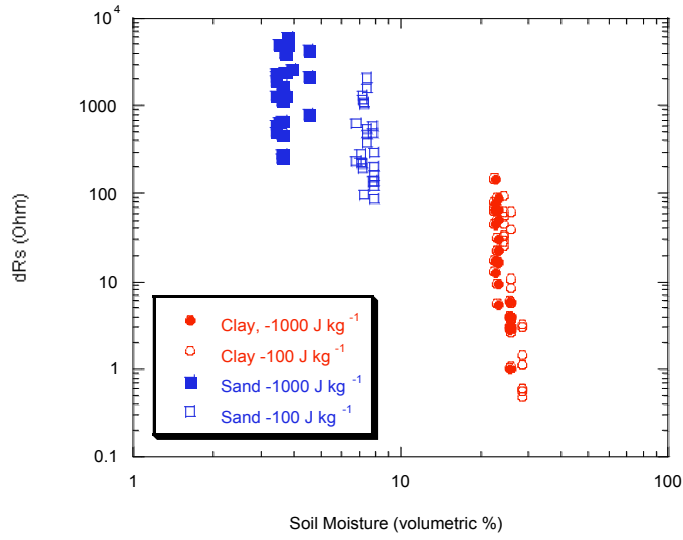

**Figure 3.** Dependence of resistance fluctuations on soil moisture at electrode distance *d*=3 cm from the wall.

In Figure 3, the impact of volumetric soil moisture on soil resistance fluctuation is shown. The scattering of the data is relatively large because of the small diameter of the electrode (3 mm). This scattering indicates the sensitivity of the method against local inhomogeneities. The scattering of data would be decreased by selecting an electrode with a wide and thin blade.

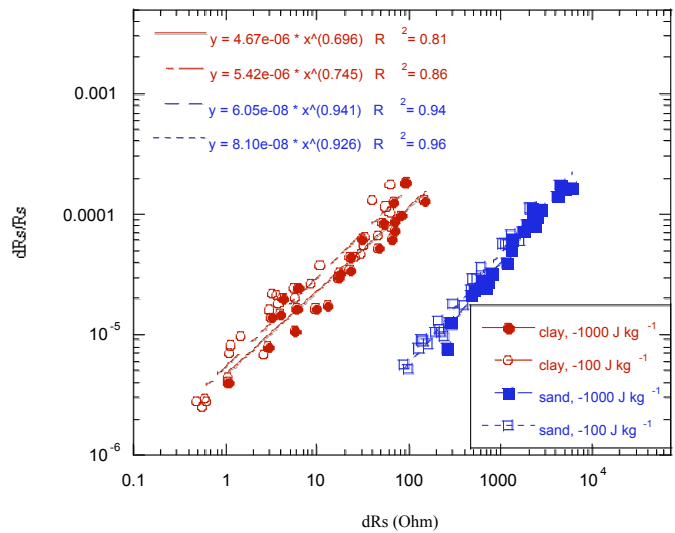

**Figure 4.** Scaling plot of different soil samples at electrode distance *d*=3cm from the wall.

The scaling plot in Figure 4 indicates strong correlations between the values of normalized resistance fluctuation and resistance fluctuation. The resistance fluctuation depends upon soil resistance (Eq. 5-6); however, the normalized resistance fluctuation does not, according to Eq. 8. Since the *dRs* and *dRs/Rs* are correlated, *dRs/Rs* provides additional information that can be helpful in measuring soil structure (porosity) independently of soil salinity. This advance provides the opportunity to develop a sensor that may have a more robust empirical calibration.

## 3. Summary

We have proposed and demonstrated a new method which is testing the mechano-electrical transport properties of soils. The normalized fluctuations show that structural electrical connectivity properties are sensitive to vibration. This information can provide additional information to moisture, texture and salinity status of a given soil measured with traditional bulk electrical conductivity measurements.

The method was briefly demonstrated with cylindrical stainless steel electrodes on clayey and sandy soils. Scattering of the data would be decreased by selecting a wide and thin blade-shape electrode.

## Acknowledgments

Appreciation is expressed to Dr. Kevin McInnes (Depart. of Soil and Crop Science, TAMU) for his advice in compaction of soil samples. The research was supported by Texas Agricultural Experiment Station. The content of this paper is the subject of a Texas A&M University patent disclosure submitted on October 31, 2005.

## References

[1] Rhoades, J. D., P. A. C. Raats and R. J. Prather, *Effects of liquid-phase electrical conductivity, water content, and surface conductivity on bulk soil electrical conductivity, Soil Sci. Soc. Am. J.* 40 (1976) 651-655.

[2] Dean, T.J., J.P. Bell and A.J.B. Baty, *Soil moisture measurement by an improved capacitance technique, Part I. Sensor design and performance. J. Hydrology* 93 (1987) 67-78.

[3] Bell, J.P., T.J. Dean, and M.G. Hodnett, *Soil moisture measurement by an improved capacitance technique, Part II. Field techniques, evaluation and calibration. J. Hydrology* 93 (1987) 79-90.

[4] Campbell, C.S., G.S. Campbell, D.R. Cobos and B. Teare, *Mitigating the effects of electrical conductivity, soil texture, and temperature on a low-cost soil moisture sensor*, In: 18[th] World Congress of Soil Science, July 9-15, 2006, Philadelphia, Pennsylvania, USA, Abstracts, p.506.

[5] Zhou, Q,Y., J. Shimada, and A. Sato, *Three-dimensional spatial and temporal monitoring of soil water content using electrical resistivity tomography*, Water Resour. Res. 37 (2001) 273-285.

[6] Michot, D. *Spatial and temporal monitoring of water flow using 2D electrical resistivity tomography in a cultivated soil: A decimeter scale study.* In: 18[th] World Congress of Soil Science, July 9-15, 2006, Philadelphia, Pennsylvania, USA, Abstracts, p.194.

[7] Corwin, D. and S.M. Lesch, *Application of soil electrical conductivity to precision agriculture: Theory, principles, and guidelines*, Agronomy Journal 95(3) (2003) 455-471.

[8] Campbell, G. S. and J. M. Norman, *An Introduction to Environmental Biophysics*, 2[nd] ed. Springer-Verlag, New York (1998) p.56.